\documentclass[10pt,letterpaper,twocolumn]{article} %% two column, final layout

\usepackage{ol2}
\usepackage[draft]{hyperref}
\usepackage{amsmath}

\begin{document}

\twocolumn[ %% activate for two-column option

\title{Effective magnetic fields for photons in waveguide and coupled resonator lattices}

%% For REVTeX it is possible to automate superscript and e-mail callouts with the superscriptaddress option; see REVTeX4 documentation.

\author{Stefano Longhi}

\address{Dipartimento di Fisica, Politecnico di Milano and Istituto di Fotonica e Nanotecnologie del Consiglio Nazionale delle Ricerche, Piazza L. da Vinci 32, I-20133 Milano, Italy (stefano.longhi@polimi.it)}

\begin{abstract}
A method to realize effective magnetic fields for photons in square lattices of coupled optical waveguides or resonators is suggested, which is inspired by an optical analogue of photon-assisted tunneling of atom optics. It is shown that an artificial magnetic field can be achieved by application of an index gradient and periodic lumped phase shifts or modulation of the propagation constants/resonances, without the need to modulate the coupling strength.  
\end{abstract}


%230.7370 Waveguides
%350.7420   Waves
%000.1600
%Classical and quantum physics

 ] %% activate for two-column option

\noindent 
Magnetism in artificial photonic structures has attracted a huge interest in the past few years with 
a promise of wide impact in optical sciences \cite{M1,M2,M3,M4,M5,M6,M7,M8,M9,M10}. Photons subjected to artificial gauge fields can
realize phenomena usually found in the domain of electronic systems, such as 
 the famous Hofstadter fractal energy bands and quantum Hall effect \cite{Hol,Moire}, or can behave like electrons in topological insulators \cite{M3,M4,M5,M8,M9,M11}.
From an applied viewpoint, a gauge transformation for the photon wavefunction can be exploited to construct 
 an optical isolator, as suggested in Ref.\cite{Refere2}. In two-dimensional (2D) structures, non-Abelian gauge fields yield topologically-protected unidirectional edge modes \cite{M3,M4,M5,M8,M9,M11}. Such states propagate immune of disorder, and thus provide a robust way of transporting light. 
So far several schemes have been proposed to realize strong artificial magnetic fields in synthetic 2D photonic structures at small length scales. 
An effective magnetic field for photons in waveguide lattices was suggested in Ref.\cite{M12}, where the magnetic force is induced by a longitudinal twist of the structure. However, the magnetic force occurs in tandem with a centrifugal potential. More recently, pseudomagnetic fields and photonic Landau levels have been proposed and experimentally demonstrated in strained honeycomb photonic lattices \cite{M9}, whereas effective
magnetic fields in a square superlattice of resonators has been suggested in Ref.\cite{M8} by harmonic modulation of the coupling constants among the resonators. Temporal modulation of the coupling between two resonators, however, requires the introduction of auxiliary  resonators and a careful design of the system\cite{M8}.\par  In this Letter we propose a method to realize artificial magnetic fields in a lattice of optical resonators or waveguides, which is inspired by an optical analogue of photon-assisted tunneling used in atom optics to create artificial magnetic fields for neutral atoms \cite{A1,A2,A3,A4}. As compared to the method of Ref.\cite{M8}, our  scheme does not require to modulate the resonator couplings,  thus avoiding the use of auxiliary cavities. 
\par
We consider light transport in
a square lattice of coupled optical resonators (as in Ref.\cite{M8}) or, similarly, spatial light propagation in a square lattice of optical waveguides \cite{M12}; see Fig.1(a). For the sake of definiteness, we will refer explicitly to latter case.
In the tight-binding and nearest-neighbor approximations, light transport in the square lattice is described by coupled-mode equations 
for the amplitudes $c_{n,m}(t)$ of the guided modes trapped in the various waveguides 
\begin{eqnarray}
i \frac{dc_{n,m}}{dt} & = &-J_x(c_{n+1,m}+c_{n-1,m})-J_y(c_{n,m+1}+c_{n,m-1}) \nonumber \\
& + & \beta_{n,m}(t) c_{n,m},
\end{eqnarray}
where $t$ is the longitudinal propagation distance, $\beta_{n,m}(t)$ is the $t$-dependent propagation constant of the waveguide at lattice site $(n,m)$ ($n,m=0, \pm1, \pm2,...$),  and $J_x$ and $J_y$ are the hopping rates between adjacent guides in the horizontal ($x$) and vertical ($y$) directions, respectively. Note that coupled-mode equations (1)  also describe temporal evolution in a coupled-resonator square lattice, provided that the spatial light propagation along the  waveguides is replaced by the temporal light evolution in the resonator system. Indeed, Eqs.(1) can be derived from the Hamiltonian
$\hat{H}(t)  =    \sum_{n,m} \beta_{n,m}(t) \hat{c}^{\dag}_{n,m} \hat{c}_{n,m} - \sum_{n,m} \left( J_x \hat{c}^{\dag}_{n,m} \hat{c}_{n+1,m} + J_y \hat{c}^{\dag}_{n,m} \hat{c}_{n,m+1}+ H.c. \right)$, which is of the form considered in Ref.\cite{M8}.  Note that, as compared to the lattice system of Ref.\cite{M8},  here we do not have a superlattice and  the hopping rates $J_{x,y}$ are not modulated in time; rather the resonance frequencies $\beta_{n,m}$  of the coupled cavities are modulated in time. This can be achieved by local dynamic modulation of the refractive index by e.g. electro-optics effect, as demonstrated for instance in Ref.\cite{Refere1} on a silicon chip. 
To introduce an effective magnetic field, we perturb $\beta_{n,m}$ from a reference value $\beta_0$ by adding a stationary gradient term along e.g. the vertical $y$ direction and a rapidly-oscillating ac term, i.e.
\begin{equation}
\beta_{n,m}(t) = \beta_0+Fm+ V_{n,m}(t)
\end{equation}  
where $F$ is the linear gradient rate and  $V_{n,m}(t+2 \pi/ \omega)=V_{n,m}(t)$ is the ac term oscillating at the spatial frequency $\omega=2 \pi / \Lambda$. A possible physical implementation of $\beta_{n,m}(t)$ for the coupled-waveguide lattice will be discussed below. The spatial frequency $\omega$ and the gradient rate $F$ are assumed to be much larger than the hopping rates $J_{x,y}$. In the absence of the ac term, i.e. for $V_{n,m}=0$, tunneling in the vertical $y$ direction is forbidden owing to the strong propagation constant mismatch between adjacent guides induced by the large index gradient. Tunneling is restored by application of the ac terms, provided that the resonance condition  
\begin{equation}
F=M \omega
\end{equation}
is satisfied for some integer $M$. This is analogous to the phenomenon of photon-assisted tunneling of cold atoms in tilted and ac-driven optical lattices \cite{A1,A2,A3}.  We further assume that the ac terms $V_{n,m}(t)$ are of the form
\begin{equation}
V_{n,m}(t)=A H(\omega t+ \phi_{n,m})
\end{equation}
where $H(x+2 \pi)=H(x)$ is a normalized periodic  function of zero mean and period $2 \pi$, $A$ is the modulation amplitude, and the phases $\phi_{n,m}$ are assumed to linearly vary with the indices $n$ and $m$, namely
\begin{equation}
\phi_{n,m}=n \sigma+m \rho
\end{equation}
where $\sigma$ and $\rho$ are two real-valued parameters, with $| \sigma, \rho| \leq \pi$. After setting $G(x)=\int_0^x dt H(t)$ and after the gauge transformation 
\begin{eqnarray}
c_{n,m}(t) & = & f_{n,m}(t)  \times \\
& \times & \exp \left [ -i \varphi_{n,m} -i \int_0^{t} dt' \beta_{n,m}(t') \right] \nonumber
\end{eqnarray}
with the phase $\varphi_{n,m}$ defined by the relations 
\begin{equation}
\varphi_{n,m}=G(\phi_{n,m})+ \frac{M}{2} \rho m (m-1) ,
\end{equation}
application of Floquet theory to Eqs.(1) yields, in the high-frequency modulation regime $\omega \gg |J_{x,y}|$, the following effective equations for the amplitudes $f_{n,m}$
 \begin{eqnarray}
 i \frac{df_{n,m}}{dt} & = & - \kappa_x f_{n+1,m}-\kappa_x^* f_{n-1,m}  \\ 
  & - & \kappa_y \exp(inM \sigma) f_{n,m+1}   -\kappa_y^* \exp(-i n M \sigma) f_{n,m-1} \nonumber
 \end{eqnarray}
where we have set
\begin{equation}
\kappa_x=\frac{J_x}{2 \pi} \int_0^{2 \pi} dx \exp \left\{ i \Gamma [G(x)-G(x+\sigma)] \right\}
\end{equation}
\begin{equation}
\kappa_y=\frac{J_y}{2 \pi} \int_0^{2 \pi} dx \exp \left\{ -iMx+i \Gamma [G(x)-G(x+\rho)] \right\}
\end{equation}
and $\Gamma= A / \omega$. Note that the effects of the ac modulation $V_{n,m}(t)$ is twofold. On the one hand, the modulation enables tunneling along the $y$ direction, with renormalized hopping rates $\kappa_x$ and $\kappa_y$ given by Eqs.(9) and (10). On the other hand, the modulation introduces the site-dependent phase shift  (Peierls phase) $nM \sigma$ in the phase of $\kappa_y$, which can not be removed by  a gauge transformation. Such a term gives rise to a non-vanishing phase for a photon that tunnels along the
borders of one plaquette. This can be interpreted as an
Aharonov-Bohm phase and therefore our square lattice serves as a model
system to study artificial magnetic fields. Indeed, Eqs.(8) are  equivalent to the equations that describe the motion of an electron with charge $e$ moving on
a square lattice in an external magnetic field $B= \hbar \sigma / Se$ \cite{Hol,A1}, where $S$ is the area of one elementary cell. In atom optics contexts \cite{A1,A2,A3,A4}, the modulation $V_{n,m}(t)$ is generally assumed to be sinusoidal, $H(x)=\cos (x) $, and the effective hopping rates are expressed in terms of Bessel functions  $\mathcal{J}_0$ and $\mathcal{J}_M$
\begin{eqnarray}
\kappa_x & = & J_x \mathcal{J}_0 \left( 2 \Gamma \sin \left( \frac{\sigma}{2} \right) \right) \\
\kappa_y & = & J_y \mathcal{J}_M \left( 2 \Gamma \sin \left( \frac{\rho}{2} \right) \right) \exp[i M (\rho-\pi)/2]. \;\;\;\;\;\;\;
\end{eqnarray}
Another interesting case in photonics is the one corresponding to a sequence of Dirac delta functions with alternating sign, namely 
$H(x)=\sum_l (-1)^l \delta(x-l \pi)$. This case is of major relevance because it effectively corresponds to application of periodic lumped phase shifts to the optical field in each waveguide, rather than to a longitudinal modulation of the propagation constants. In fact, from the coupled-mode equations (1) it readily follows that the choice $H(x)=\sum_l (-1)^l \delta(x-l \pi)$  effectively lead to the periodic application, along the longitudinal propagation direction $t$, of lumped phase shifts to the optical field in each waveguide of the array, of strength $\Gamma=A/ \omega$ and of alternating sign, at the spatial positions $t_{n,m}^{(l)}=(l \pi - \phi_{n,m}) / \omega$ ($l$ integer) defined by the phases $\phi_{n,m}$.  In this case the effective hopping rates are 
 \begin{figure}[htb]
\centerline{\includegraphics[width=8.6cm]{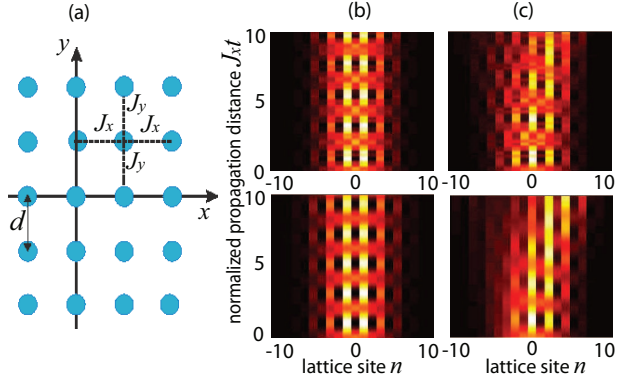}} \caption{
(Color online)  (a) Schematic of a square lattice of optical waveguides / resonators. (b) Interference fringes [evolution of $I_n(t)=\sum_{m}|c_{n,m}(t)|^2$] in a square lattice of waveguides illuminated by a broad Gaussian beam for a rational flux number $\alpha=1/2$ as obtained by numerical simulations of Eqs.(1) (upper panel) and from the effective equations (8) (lower panel). (c) Same as (b), but for an irrational flux number $\alpha=3/(2 \pi)$.  Parameter values are given in the text.}
\end{figure}
\begin{eqnarray}
\kappa_x & = & J_x \left( 1- \frac{2 |\sigma|}{\pi} \sin^2 \left( \frac{\Gamma}{2} \right) \right) \\
\kappa_y & = & \frac{4J_y}{M \pi} \sin \left( \frac{M \rho}{2} \right) \sin \left( \frac{\Gamma}{2} \right) \sin \left( \frac{M \pi}{2}-\frac{\rho}{|\rho|}\frac{\Gamma}{2} \right) \nonumber \\
& \times &  \exp[i M (\rho-\pi)/2]. 
\end{eqnarray}
 \begin{figure}[htb]
\centerline{\includegraphics[width=8.6cm]{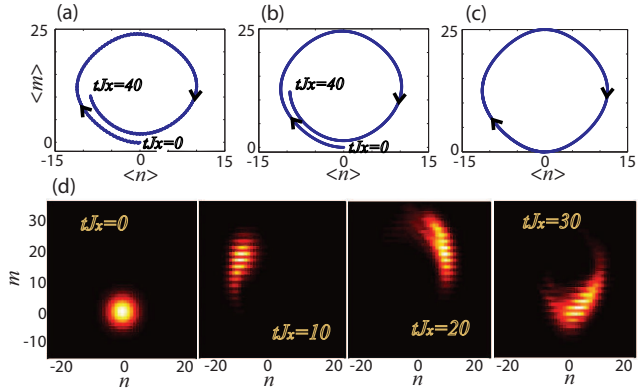}} \caption{
(Color online)  Cyclotron motion of an initial Gaussian wave packet [path described by the beam center of mass $(\langle n(t) \rangle, \langle m(t) \rangle]$ as obtained (a) from numerical simulations of Eqs.(1), (b) from the effective equations (8), and (c) from the semiclassical analysis. In (d) snapshots of the beam intensity distribution $|c_{n,m}(t)|^2$ at a few normalized propagation distances $J_xt$ are shown.  Parameter values are given in the text.}
\end{figure}
The main result of our scheme is thus that an artificial magnetic field can be realized in a square lattice of waveguides by application of a transverse index gradient and longitudinal lumped phase shifts, without the need for modulating neither the coupling rates nor the propagation constants.\par 
To show the correctness of our scheme, let us discuss two effects of the artificial magnetic field on light transport  in a waveguide lattice. The first one is the interference pattern that arises when illuminating the square lattice by a broad (plane wave) beam, which critically depends on whether the flux number  $ \alpha \equiv \sigma M / (2 \pi)$ is a rational or an irrational number \cite{Hol,A1}.  For $\sigma=0$, i.e. in the absence of the magnetic field, the square lattice has a purely continuous spectrum with a single band. For $\sigma \neq 0$, the  spectrum  sensitively depends on the ratio $\kappa_x/ \kappa_y$ and on the parameter $\alpha$, and it is related to the spectrum of the Harper equation for $\kappa_x=\kappa_y$ \cite{Hol} and of the Aubry-Andr\'{e} equation for $\kappa_y \neq \kappa_x$ \cite{Aubry}. Considering the case $\kappa_x=\kappa_y$, if $\alpha= p/q$ is a rational number the energy spectrum splits into a finite number of exactly $q$ bands, whereas if $\alpha$ is irrational the energy spectrum in continuous-singular, i.e. it breaks up into infinitely many bands \cite{Hol,A1}. The different band structure in the two cases can be visualized by monitoring the evolution of the integrated intensity light pattern in the vertical plane, namely $I_n(t)=\sum_{m} |c_{n,m}(t)|^2$, when the lattice is initially excited by a (nearly) plane wave with phase imprinting $\varphi_{n,m}$. As an example, in Figs.1(b) and (c) we plot the evolution of $I_{n}(t)$ as obtained by numerical simulations of Eqs.(1) for a sinusoidal modulation $H(x)=\cos(x)$ and for (a) a rational $\alpha=1/2$, and (b) an irrational $\alpha=3 /(2 \pi)$ flux number;  parameter values are $J_2=J_1 \equiv J$, $\omega=8 J$, $M=1$, $\Gamma=0.717$, and $\rho=\pi$, corresponding to $\kappa_x \simeq \kappa_y \simeq 0.548 J$. The initial condition has been assumed to be a broad beam with Gaussian shape
$c_{n,m} \propto \exp[-(n^2+m^2)/w^2] \exp(-i \varphi_{n,m})$ of width $w=5$, where the phases $\varphi_{n,m}$ are given by Eq.(7). The lower panels in the figure show the evolution of $I_n(t)$ as predicted by the averaged equations (8). Note that, in the rational $\alpha=p/q=1/2$ case [Fig.1(b)], interference fringes in the pattern of $I_n(t)$ are clearly visible, with spatial lattice periodicity $q=2$ along the $x$ axis, and the evolution is periodic in $t$ with a period $\Delta t \simeq 2 J$, which is determined by the cyclotron frequency $\Omega_c \sim 2 \sigma \kappa_x$. Such fringes arise because the magnetic field splits the original lattice band into $q=2$ minibands; the initial spatially-broad beam excites Bloch modes of the two minibands, and the beating between these two modes causes the periodic appearance and disappearance of the fringes. When $\alpha$ is slightly changed to assume an irrational value, $\alpha=3/(2 \pi)$ [Fig.1(c)], the visibility of the fringes is degraded and the periodicity of the pattern along $t$ is broken after few revivals. The reason thereof is that in this case the tight-binding energy band breaks into infinitely many bands \cite{Hol}. The results shown in Figs.1(b) and (c) are obtained for a sinusoidal shape of $H(x)$, however the same behavior is found for the $\delta$-Dirac case discussed above, i.e. by application of lumped phase shifts. \par
The second effect is the spiraling (cyclotron) motion of an optical beam induced by the Lorentz force. The evolution of a broad beam wave packet in the lattice is influenced by both the magnetic force and the lattice band structure, so that spiraling is generally not circular owing to the dependence of the mass on the momentum. The path followed by the beam can be derived from a semiclassical analysis \cite{M12}. After introduction of the generalized momenta $\hat{P}_n=- i \partial_n $ and $\hat{P}_m=- i \partial_m+n \sigma$, the following equations for mean values can be readily obtained from Eqs.(8)
\begin{eqnarray}
\frac{d \langle n \rangle}{dt}= 2 \kappa_x \langle \sin \hat{P}_n \rangle , \;\; \frac{d \langle m \rangle}{dt}= 2 \kappa_y \langle \sin \hat{P}_m \rangle \;\;\; \\
\frac{d \langle \hat{P}_n \rangle}{dt}= -2 \kappa_y \sigma \langle \sin \hat{P}_m \rangle , \;\; \frac{d \langle \hat{P}_m \rangle}{dt}= 2 \kappa_x \sigma \langle \sin \hat{P}_n \rangle \;\;\;
\end{eqnarray}
where we assumed $\kappa_{x,y}$ real-valued without loss of generality and where $\langle ... \rangle$ denotes the mean value (for instance, $\langle n \rangle \equiv \sum_{n,m} n |f_{n,m}|^2 / \sum_{n,m} |f_{n,m}|^2$). In the semiclassical limit one can assume $ \langle \sin \hat{P}_{n,m} \rangle \simeq  \sin ( \langle \hat{P}_{n,m} \rangle )$, so that Eqs.(15) and (16) provide a nonlinear system of equations that describe the evolution of the mean position $\langle n \rangle$,  $\langle m \rangle$ and generalized momenta $ \langle \hat{P}_{n} \rangle $,  $\langle \hat{P}_{m} \rangle$. As an example, in Fig.2(a) we show the path followed an initial Gaussian beam that excites a square waveguide lattice as obtained by numerical simulations of Eqs.(1) for a sinusoidal modulation of $V_{n,m}$ and for parameter values $J_y / J_x=2$, $\Gamma=0.9$, $\rho=\pi$,  $\sigma=-\pi/25$, corresponding to $k_x \simeq J_x$ and $\kappa_y \simeq 1.16 J_x$.  The initial condition is a broad Gaussian beam with an initial transverse velocity $p$ along the horizontal $x$ axis and of size $w$, namely  $c_{n,m} \propto \exp[-(n^2+m^2)/w^2-ipn] \exp(-i \varphi_{n,m})$ with $w=5$, $p=\pi /2$ and where the phases $\varphi_{n,m}$ are given by Eq.(7). Figures 2(b) and (c) show, for comparison, the beam paths as obtained from the effective lattice equations (8) and from the semiclassical analysis [Eqs.(15) and (16)], respectively. Snapshots of the beam intensity distribution $|c_{n,m}(t)|^2$ at a few propagation distances $t$ are depicted in Fig.2(d),  showing beam deformation as the light bend along the cyclotron path. \par 
Finally, let us briefly discuss practical implementations of our proposal. A first possibility, which fits the status of current experimental technology, is to manufacture a square lattice of circularly-bent waveguides using fs laser writing in fused silica \cite{Sza}. The waveguide bending along the $y$-axis introduces a gradient index with rate  $F=2 \pi n_s d / (\lambda R) $, where $R$ is the bending radius of curvature,  $d$ is the vertical spacing of waveguides, $\lambda$ is the photon wavelength, and $n_s$ is the substrate refractive index \cite{LonghiLPR}. A longitudinal periodic modulation $V_{n,m}(t)$ of the propagation constants can be realized  by periodically varying the refractive index contrast $\delta n$ and/or the size of the waveguide channel, as demonstrated in Ref. \cite{Sza}. Alternatively,  a periodic sequence of lumped phase shifts along the propagation direction can be realized by short waveguide segmentation \cite{APL}. To get an idea of physical parameters corresponding to the simulations of Figs.1(b) and (c), let us consider a lattice period $d=19 \; \mu$m, a probing light beam at  $\lambda=633$ nm with typical coupling rate $J_x \sim J_y  \sim 1 \; {\rm cm}^{-1}$. The bending radius of curvature of the guides then turns out to be $R \simeq 34$ cm, the spatial modulation period $\Lambda$ of the propagation constant is $\Lambda = 2 \pi / \omega \simeq 7.85$ mm, the modulation amplitude is $A = \Gamma \omega \simeq 5.74 \; {\rm cm}^{-1}$ (corresponding to an index change $\delta n \sim \lambda A/ 2 \pi \simeq 5.78 \times 10^{-5}$), and the total sample length is $L=10/J_x=10$ cm. A second possible implementation, which can be foreseen with the current status of technology,  is based on a coupled-resonator system realized  in a photonic crystal, as in Ref.\cite{M8}. The static resonance  frequencies of the microcavities should be tailored, either electro-optically or by appropriate design of the defects in the crystal,  to generate the gradient term along the $y$ direction, whereas  harmonic modulation in time with controlled phases 
can be obtained as discussed in Refs.\cite{M8,Refere1}. Note that, as compared to the design of Ref.\cite{M8}, here we do not need auxiliary resonators. \par
In conclusion, a method for realizing artificial magnetic fields for photons in a square lattice of coupled optical waveguides or resonators has been proposed, which is inspired by an optical analogue of photon-assisted tunneling of atom optics. Our results could be of relevance for both applied and basic aspects, for example for the design of 2D photonic devices with edge modes propagating immune of disorder or for  the realization of photonic simulators of electronic systems in magnetic fields \cite{Hol,Moire,dHarper}, in which the use of photons rather than electrons avoids decoherence and many-body effects.

\newpage

\footnotesize {\bf References with full titles}\\
\\
1. S. Raghu and F.D.M. Haldane, "Analogs of quantum-Hall-effect edge states in
photonic crystals", Phys. Rev. A {\bf 78}, 033834 (2008).\\
2. F.D.M. Haldane and S. Raghu, "Possible realization of directional optical
waveguides in photonic crystals with broken time-reversal symmetry",
Phys. Rev. Lett. {\bf 100}, 013904 (2008).\\
3. Z. Wang, Y. Chong, J.D. Joannopoulos,  and M. Soljacic,  "Reflection-free
one-way edge modes in a gyromagnetic photonic crystal", Phys. Rev. Lett.
{\bf 100}, 013905 (2008).\\
4. Z. Wang, Y. Chong, J.D. Joannopoulos, and M. Soljacic, "Observation of
unidirectional backscattering-immune topological electromagnetic states",
Nature {\bf 461}, 772Ð775 (2009).\\
5.  M. Hafezi, E.A. Demler, M.D. Lukin, and J.M. Taylor, "Robust optical delay lines
with topological protection", Nature Phys. {\bf 7}, 907Ð912 (2011).\\
6. R. O. Umucallar and I. Carusotto, "Artificial gauge field for photons in coupled
cavity arrays", Phys. Rev. A {\bf 84}, 043804 (2011).\\
7. K. Fang, Z. Yu, and S. Fan, "Photonic Aharonov-Bohm effect based on dynamic
modulation", Phys. Rev. Lett. {\bf 108}, 153901 (2012).\\
8. K. Fang, Z. Yu, and S. Fan, "Realizing effective magnetic field for photons by
controlling the phase of dynamic modulation", Nature Photon. {\bf 6}, 782-787 (2012).\\
9.  M.C. Rechtsman, J.M. Zeuner, A. T\"{u}nnermann, S. Nolte,	
M. Segev, and A. Szameit, "Strain-induced pseudomagnetic field and photonic Landau levels in dielectric structures", Nature Photon. {\bf 7}, 153Ð158 (2013).\\
10. L. Lu, L. Fu, J.D. Joannopoulos, and M. Soljacic, "Weyl points and line nodes in gyroid photonic crystals",  Nature Photon. {\bf 7}, 294-299 (2013).\\ 
11. D.R. Hofstadter,  " Energy levels and wave functions of Bloch electrons in rational and irrational magnetic fields", Phys. Rev. B {\bf 14}, 2239 (1976).\\
12.  C.R. Dean,	 L. Wang,	P. Maher,	C. Forsythe, F. Ghahari, Y. Gao, J. Katoch, M. Ishigami, P. Moon, M. Koshino, T. Taniguchi, K. Watanabe, K. L. Shepard,	 J. Hone, and P. Kim, "HofstadterÕs butterfly and the fractal quantum Hall effect in moirŽ superlattices",  Nature (2013); doi: 10.1038/nature12186.\\
13. A.B. Khanikaev,	S. Hossein Mousavi, W.-K. Tse, M. Kargarian, A.H. MacDonald, and G. Shvets, "Photonic topological insulators", Nature Mat. {\bf 12}, 233Ð239 (2013).\\
14. K. Fang,  Z. Yu, and S. Fan, "Photonic Aharonov-Bohm effect based on dynamic modulation", Phys. Rev. Lett. {\bf 108}, 153901 (2012).\\
15. S. Longhi, "Bloch dynamics of light waves in helical optical waveguide arrays", Phys. Rev. B {\bf 76}, 19511 (2007).\\
16. D. Jaksch and P. Zoller, "Creation of effective magnetic fields in optical lattices: the Hofstadter butterfly for cold neutral atoms", New J. Phys. {\bf 5}, 56 (2003).\\
17. A. R. Kolovsky,  "Creating artificial magnetic fields for cold atoms by photon-assisted tunneling", EPL {\bf 93}, 20003 (2011).\\ 
18. M. Aidelsburger, M. Atala, S. Nascimbene, S. Trotzky, Y.-A. Chen, and I. Bloch, "Experimental Realization of Strong Effective Magnetic Fields in an Optical Lattice", Phys. Rev. Lett. {\bf 107}, 255301 (2011).\\
19. A. Bermudez, T. Schaetz, and D. Porras, "Photon-assisted-tunneling toolbox for quantum simulations in ion traps", New J. Phys. {\bf 14}, 053049 (2011).\\
20. H. Lira, Z. Yu, S. Fan, and M. Lipson, "Electrically driven nonreciprocity induced by interband photonic transition on a silicon chip", Phys. Rev. Lett.
{\bf 109}, 033901 (2012).\\
21. S. Aubry and G. Andr\'{e}, "Analyticity breaking and Anderson localization in incommensurate lattices",  Ann. Israel. Phys. Soc. {\bf 3}, 133 (1980).\\
22. A. Szameit, Y. V. Kartashov, F. Dreisow, M. Heinrich, T. Pertsch, S. Nolte, A. T\"{u}nnermann, V. A. Vysloukh, F. Lederer, and L. Torner, ""Inhibition of Light Tunneling in Waveguide Arrays", Phys. Rev. Lett. {\bf 102}, 153901 (2009).\\
23. S. Longhi, "Quantum-optical analogies using photonic
structures", Laser and Photon. Rev. {\bf 3}, 243-261 (2009).\\
24.  A. Szameit, F. Dreisow, M. Heinrich, T. Pertsch, S. Nolte, A. T\"{u}nnermann, E. Suran, F. Louradour, A. Barthelemy, and
S. Longhi, "Image reconstruction in segmented femtosecond laser-written waveguide arrays", Appl. Phys. Lett. {\bf 93}, 181109 (2008).\\
25. A.R. Kolovsky and G. Mantica, "The driven Harper model", Phys. Rev. B {\bf 86}, 054306 (2012).

\end{document}